\documentclass[10pt,A4paper,amsmath,amssymb,aps,etoolbox,floatfix,letterpaper,
noeprint,nofootinbib,
noshowpacs, 
preprintnumbers,reprint,showkeys,
twocolumn]
{revtex4}

\newcommand{\beq}{\begin{equation}}

\newcommand{\eeq}{\end{equation}}
\newcommand{\ds}{\displaystyle}

\usepackage{array}
\usepackage{bm}       
\usepackage{color}
\usepackage{dcolumn}  
\usepackage{epsfig}
\usepackage{float}
\usepackage{gensymb}
\usepackage{graphicx} 
\usepackage{hyperref}
\usepackage{longtable}
\usepackage{mathtools}
\usepackage{natbib}
\usepackage{pdflscape}
\usepackage{sidecap}
\usepackage[normalem]{ulem}

\graphicspath{%
    {converted_graphics/}
    {/}
}
\begin{document}

\title[SNeIa and photon frequency shift]{Supernovae and photon frequency shift induced by the Standard-Model Extension}

\author
{
Alessandro D.A.M. Spallicci\textsuperscript{a,b,c,d,e,h}\footnote{spallicci@cnrs-orleans.fr, 
http://wwwperso.lpc2e.cnrs.fr/$\sim$spallicci/\\
- French Chair UERJ and Visiting Professor CBPF.\\
- Erasmus Professor UNINA.}}
\author {Fabio Ragosta,\textsuperscript{e,f,g}\footnote{fabio.ragosta@inaf.it}\\
Jos\'e A. Helay\"el-Neto\textsuperscript{h}\footnote{josehelayel@gmail.com 
}}
\author{Mart\'in L\'opez-Corredoira\textsuperscript{i,j}\footnote{fuego.templado@gmail.com 
}}

\affiliation{
\mbox{\textsuperscript{a}Universit\'e d'Orl\'eans, Observatoire des Sciences de l'Univers en r\'egion Centre (OSUC) UMS 3116} \\
\mbox{1A rue de la F\'{e}rollerie, 45071 Orl\'{e}ans, France}
\vskip1pt
\mbox{\textsuperscript{b}Universit\'e d'Orl\'eans, Collegium Sciences et Techniques (CoST), P\^ole de Physique}\\
\mbox{Rue de Chartres, 45100 Orl\'{e}ans, France} 
\vskip1pt
\mbox{\textsuperscript{c}Centre National de la Recherche Scientifique (CNRS)}\\
\mbox{Laboratoire de Physique et Chimie de l'Environnement et de l'Espace (LPC2E) UMR 7328}\\
\mbox {Campus CNRS, 3A Avenue de la Recherche Scientifique, 45071 Orl\'eans, France}
\vskip1pt
\mbox{\textsuperscript{d}Universidade do Estado do Rio de Janeiro (UERJ), Instituto de F\'{i}sica, Departamento de F\'{i}sica Te\'{o}rica}\\
\mbox{Rua S\~ao Francisco Xavier 524, Maracan\~a, Rio de Janeiro, 20550-013, Brasil}
\vskip1pt
\mbox{\textsuperscript{e}Universit\`a degli Studi di Napoli, Federico II (UNINA), Dipartimento di Fisica Ettore Pancini}\\
\mbox{Via Cinthia 9, 80126 Napoli, Italia}
\vskip1pt
\mbox{\textsuperscript{f}Istituto Nazionale di Astrofisica (INAF), Osservatorio Astronomico di Capodimonte}\\
\mbox{via Moiariello 16, 80131 Napoli, Italia} 
\vskip1pt
\mbox{\textsuperscript{g}Istituto Nazionale di Fisica Nucleare (INFN), Sezione di Napoli}\\
\mbox{via Cinthia 9, 80126 Napoli, Italia} 
\vskip1pt
\mbox{\textsuperscript{h}Centro Brasileiro de Pesquisas F\'{\i}sicas (CBPF), Departamento de Astrof\'{\i}sica, Cosmologia e Intera\c{c}\~{o}es Fundamentais (COSMO)}\\
\mbox {Rua Xavier Sigaud 150, 22290-180 Urca, Rio de Janeiro, RJ, Brasil}
\vskip1pt
\mbox{\textsuperscript{i}Instituto de Astrof{\'i}sica de Canarias (IAC)}\\
\mbox {C/ V{\' i}a L{\'a}ctea, s/n, 38205 La Laguna (Tenerife), Espa\~na}
\vskip1pt
\mbox{\textsuperscript{j}Universidad de La Laguna, Facultad de Ciencias, Secci{\' o}n de F{\'i}sica, Departamento de Astrof{\'i}sica}\\
\mbox{Avenida Astrof{\'i}sico Francisco S{\'a}nchez, s/n, Apartado 456, 38200 San Crist{\'o}bal de La Laguna (Tenerife), Espa\~na}
}

\date{12 December 2019}

\begin{abstract}
We revisit for 714 SNeIa the discrepancy between the red-shift associated to the distance modulus $\mu$ and the spectroscopic red-shift. Previous work has shown that the total red-shift $z$ might be a combination of the expansion red-shift $z_{\rm C}$ and of a static, blue or red shift $z_{\rm LSV}(r)$, $r$ being the comoving distance. The latter is due to the energy non-conservation of the photon propagating 
through Electro-Magnetic (EM) background fields (host galaxy, intergalactic and Milky Way), under Lorentz(-Poincar\'e) Symmetry Violation (LSV), associated to the Standard-Model Extension (SME). The non-conservation stems from the vacuum expectation value of the vector and tensor LSV fields. For zero radiation $\Omega_{\rm rad}$ and curvature $\Omega_{\rm k}$ densities, and matter density $\Omega_m = 0.28$, the SN1a positions in the ($\mu$, z) plan are recovered according to the different strengths, orientations, alignments and space-time dependencies of the EM fields and LSV components. The LSV vacuum energy may be thus tantamount to $\Omega_\Lambda \simeq 0.7$, but unrelated to an accelerated expansion. We present models with red or blue-shifts $z_{\rm LSV}$, below $10\%$ of $z$. The $\nu$ frequency variation is below $10^{-19} \Delta \nu/\nu$ per m. 
\end{abstract}


\maketitle

{\it Introduction.} 
At the infinitesimal scale, the Standard-Model (SM) unifies three of the four fundamental forces: Electro-Magnetic (EM), weak, and strong interactions. The SM guarantees Lorentz(-Poincar\'e) Symmetry (LoSy), stating that experimental results are independent of the time, position or velocity of the observer. 
Despite its remarkable successes, the SM does not incorporate the particles corresponding to a, yet to be experimentally found, dark universe. Indeed, it does not account for the accelerating expansion of the universe (dark energy) nor has a viable dark matter particle. Furthermore, the photon is the only free massless particle in the SM, after the attribution of some mass to the neutrino. 
Thus, a theory  going beyond the SM has been put forward: the SM Extension (SME) \cite{colladaykostelecky1997,colladaykostelecky1998}. 
The SM is assumed LoSy invariant up to certain energy scales beyond which a LoSy Violation (LSV) might occur. The SME 
 allows testing the low-energy manifestations of the LSV. 

{\it Light propagation in the SME.} 
In \cite{bodshnsp2017,bodshnsp2018} the foundations of the analysis of light propagation in the SME were laid. Going beyond the SM, we encounter a massive photon, compatible with the actual upper limit \cite{tanabashietal2018} of $10^{-54}$ kg, but as pointed out 
in \cite{retinospalliccivaivads2016}, this limit arises chiefly from modelling rather than measurements.  
The effective, and frame dependent, mass is proportional to the value of the LSV parameters and, conversely to the de Broglie \cite{db23,db40}-Proca \cite{proca-1937} formalism, the SME massive photon is gauge-invariant \cite{bodshnsp2017,bodshnsp2018}. 
Moreover, in specific conditions, sub- and super-luminal velocities, imaginary and complex frequencies, birefringence appear, and, evidently in any condition, LSV anisotropy and inhomogeneity are present. The effects of the latter two are not dominant in the regimes we explored. The massive photon group velocity differs from $c$ of a quantity proportional to the inverse of the frequency squared \cite{db40}. 
Incidentally, such dependence has been analysed recently with the signals from Fast Radio Bursts (FRBs) \cite{boelmasasgsp2016,wuetal2016b,boelmasasgsp2017,bebosp2017,shao-zang-2017,xing-etal-2019}. Most remarkably, the energy-momentum tensor of a light-wave is not conserved in vacuum when light crosses an EM background field \cite{bodshnsp2018}.

Non-conservation occurs also when the EM background field and the LSV perturbations are independent from space-time coordinates, but interestingly, for a space-time dependent LSV, the presence of an EM field is not necessary \cite{helayel-spallicci-2019}. Indeed, the photon exchanges energy with the LSV and EM fields. 
An estimate of the energy change that light would undergo was given \cite{helayel-spallicci-2019}. 
The wave-particle correspondence, even for a single photon \cite{aspect-grangier-1987}, leads to consider that the light-wave energy non-conservation is translated into photon energy variation and thereby into a red or a blue shift. The energy variations, if losses, would translate into frequency damping.  

The LoSy breaking factors are represented by a $k^{\rm AF}_{\alpha}$ four-vector when the handedness of the Charge conjugation-Parity-Time reversal (CPT) symmetry is odd and by a $k_{\rm F}^{\alpha\nu\rho\sigma}$ tensor when even. 
The physical reasons inducing non-conservation notwithstanding constant EM background and LSV breaking vector (the breaking tensor appears either under a derivative or coupled to a derivative of the EM background) lie in the Carroll-Field-Jackiw action \cite{cafija90} for the CPT odd handedness. The action contains a contribution such that even if the EM background (and the LSV perturbation) is constant, the corresponding four-potential is not. Thereby, there is an explicit $x^\alpha$ coordinate dependence at the level of the Lagrangian. This determines a source of energy-momentum non-conservation, according to the Noether theorem. Otherwise put, there is an exchange of energy-momentum between the photon and the EM background.

{\it The LSV frequency shift.}
For $f_{\alpha\nu}$ representing the photon field,  $a_\nu$ the four-potential, $F_{\alpha\nu}$ the EM background field, $j^{\nu}$ an  external current and the symbol * the dual field, the energy-momentum density flux 
$\theta_{\ 0}^{\alpha}$ variation is given by \cite{bodshnsp2018} (in SI units $\mu_0 = 4\pi \times 10^{-7}$ NA$^{-2}$)

\vspace*{-0.2cm} 

\begin{eqnarray}
\partial_{\alpha}\theta_{\ 0}^{\alpha} & = j^{i}f_{i 0}
- {\ds \frac{1}{\mu_0}} \left [\left(\partial_{\alpha}F^{\alpha i}\right)f_{i 0}
+ k^{\rm AF}_{\alpha}\ ^{*}F^{\alpha i}f_{i 0} + \right.\nonumber \\
 & \frac{1}{2}\left(\partial_{\alpha}k^{\rm AF}_{0}\right)\ ^{*}f^{\alpha\nu}a_{\nu}
-  \frac{1}{4}\left(\partial_{0}k_{\rm F}^{\alpha i\kappa\lambda}\right)f_{\alpha\nu}f_{\kappa\lambda} + \nonumber \\
 & \left. \left(\partial_{\alpha}k_{\rm F}^{\alpha i\kappa\lambda}\right)F_{\kappa\lambda}f_{i 0}
+ k_{\rm F}^{\alpha i \kappa\lambda}\left(\partial_{\alpha}F_{\kappa\lambda}\right)f_{i 0}\right]~.
\label{pemt-nc-0}
\end{eqnarray}

\begin{table}
\caption{\footnotesize LSV shift types. In the first, the frequency variation is proportional to the instantaneous frequency and to the distance; in the second to the emitted frequency and the distance; in the third only to the distance (we don't consider a fourth option proportional to the observed frequency and the distance). $k_{1,2}$ have the dimensions of Mpc$^{-1}$, $k_3$ of Mpc$^{-1}$s$^{-1}$. The positiveness of the distance $r$ constraints $z_{\rm LSV/1} > - 1$ for $k_1<0$, and $- 1 < z_{\rm LSV/1} < 0 $ for $k_1 >0$.}
\label{tab1}
\newcolumntype{C}[1]{>{\centering\arraybackslash}m{#1}}
\setlength\extrarowheight{3pt}
\begin{tabular}
{ 
| C{1cm} 
| C{2.4cm} 
| C{2.4cm} 
| C{2.4cm} 
|
} 
\hline
Type
& 1
& 2 
& 3 
\\ \hline
$\Delta \nu $  
& $k_1 \nu dr$
& $k_2 \nu_{\rm e} dr$
& $k_3 dr$ 
\\ \hline
$\nu_{\rm o}$ 
& $\nu_{\rm e}\exp^{k_1 r}$  
& $\nu_{\rm e}(1 + k_2 r)$         
& $\nu_{\rm e}+ k_3 r$ 
\\ \hline
$z_{\rm LSV}$ 
& $\exp^{- k_1 r} - 1$                      
& $-{\ds \frac{k_2 r}{1 + k_2 r}}$  
& $-{\ds \frac{k_3 r}{\nu_e + k_3 r}}$ 
\\ \hline
\end{tabular}
\vskip5pt
\end{table}

The dimensions are Jm$^{-4}$. In absence of an external current and of a large scale electric field, but considering only the constant components of the galactic and inter-galactic magnetic fields, and finally neglecting at optical frequencies the contribution of $k_{\rm F}^{\alpha i \kappa\lambda}$, an estimate was given through a simplified expression along the line of sight observer-source \cite{helayel-spallicci-2019}

\vspace*{-0.2cm} 

\beq
\partial_{\alpha}\theta_{\ 0}^{\alpha} \approx  
-\frac{1}{\mu_0} k^{\rm AF}_{0}\ ^{*}F^{0 i}f_{i 0}~.
\label{pemt-nc-0-new-expl}
\eeq
The $^{*}F^{0 i}$ represents the magnetic components of the dual EM background tensor field, $f_{i 0}$ the electric components of the EM photon tensor field, and $k^{\rm AF}_0$ is the time component of the breaking vector. Assuming that the energy variation of a light-wave corresponds to a frequency shift for a photon, we converted the energy variation of Eq. (\ref{pemt-nc-0-new-expl}) into a $\Delta\nu$, being $\nu$ the photon frequency.  

After some steps, an exact expression for $|\Delta\nu|$ due to LSV was given \cite{helayel-spallicci-2019}. Herein, we reproduce it qualitatively  

\vspace*{-0.2cm} 

\beq
|\Delta\nu|_{\rm LSV} \propto B~f_{i 0}~t_{\rm LB}~ k^{\rm AF}_0~{\varrho} ~,
\label{estimate3new}
\eeq
where ${\varrho}$, is a positive arbitrary safe margin, taking into account that the magnetic fields (of the galaxy of the source, inter-galactic, and the Milky Way) estimated at $B = 5 \times 10^{-10} - 5 \times 10^{-9}$ T each, and crossed by light from the source to us in a look-back time $t_{\rm LB}$, have likely different orientations and partly compensate their effects on the wave  electric field $f_{i 0}$. We have not considered any magnetic field at the source. 

In the following, we refer to a general $z_{\rm LSV}$ given by all the terms in Eq. (\ref{pemt-nc-0}).  

{\it The LSV as vacuum energy.}
According to \cite{kosteleckysamuel1989b}, the LoSy breaking four-vector, $k_{\rm AF}$, and the rank-four tensor, $k_{\rm F}$, correspond, respectively, to the vacuum condensation of a vector and a tensor field in the context of string models. They describe part of the vacuum structure, which appears in the form of space-time anisotropies. Therefore, their presence at the right-hand side of Eq. 
(\ref{pemt-nc-0}) reveals that vacuum effects are responsible for the energy variation of light waves, which, in turn, correspond to a photon frequency shift. Otherwise put, vacuum anisotropies are the true responsible for changing the frequency of light emitted by astrophysical structures.

{\it Superposing the shifts.} 
We recall that $z = \Delta \nu/\nu_o$ where $\Delta \nu = \nu_{\rm e} - \nu_o$ is the difference between the observed $\nu_o$ and emitted 
$\nu_{\rm e}$ frequencies, or else $z = \Delta \lambda/\lambda_{\rm e}$ for the wavelengths. 

We pose the following conjecture. Expansion causes $\lambda_{\rm e}$ to stretch to $\lambda_{\rm c}$, that is $\lambda_{\rm c} = (1+z_{\rm C})\lambda_{\rm e}$. The wavelength $\lambda_{\rm c}$ could be further stretched or shrunk for the LSV shift to $\lambda_{\rm o} = (1+z_{\rm LSV})\lambda_{\rm c}  = (1+z_{\rm LSV}) (1+z_{\rm C}) \lambda_{\rm e}$. But since $\lambda_{\rm o} = (1+z)\lambda_{\rm e}$, we have 
$1 +  z = (1+z_{\rm C}) (1+z_{\rm LSV})$; thus
\vspace*{-0.2cm} 

\begin{eqnarray}
z = z_{\rm C} + z_{\rm LSV}  + z_{\rm C}z_{\rm LSV}~. 
\label{newz}
\end{eqnarray}

The second order is not negligible for larger $z_{\rm C}$. 

{\it Behaviour of the LSV shift with distance.\label{zmodels}}
We model the behaviour of $z_{\rm LSV}$ with distances in three different ways, Tab. \ref{tab1}. Since $z_{\rm C}$ stems from  the universe expansion, we estimate $z_{\rm LSV}$ on the basis of the comoving distance. The $k_i$ parameters, proportional to the Lema{\^i}tre-Hubble-Humason constant $H$, herein $70$ km/s per Mpc, take either positive (frequency increase) or negative (frequency decrease) values. 

{\it The LSV frequency shift from the SNIa data.}
A greater than expected SNIa distance modulus for a given red-shift led to consider a cosmological constant or dark energy \cite{riess-etal-1998, perlmutter-etal-1999}. Later, suppositions of alternatives theories of gravity, see \cite{lopezcorredoira-2017,lopezcorredoira-2018} for reviews, and of photons oscillating into axions producing a dimming effect for SNeIa 
 \cite{csaki-kaloper-terning-2002a,csaki-kaloper-terning-2002b} were made.  
It must be recalled that SNIa data reliability is still debated \cite{vishwakarma-narlikar-2010,nielsen-guffanti-sarkar-2016,rubin-hayden-2016,dam-heinesen-wiltshire-2017,haridasu-lukovic-dagostino-vittorio-2017,velten-gomes-busti-2018,colin-etal-2019}. 
We challenge the meaning of the red-shift uniquely explained by expansion and conceive a complementary point of view.

For appreciating $d_{\rm L}$, as function of $z_{\rm C}$, we have computed for 
for $\Omega_{\rm rad} = \Omega_{\rm k} = \Omega_\Lambda = 0$ 

\beq
d_{\rm L}=\frac{2c}{H \Omega_{\rm m}^2}\left[2 - \Omega_{\rm m} (1-z_{\rm C}) - (2 - \Omega_{\rm m})\sqrt{1 + \Omega_{\rm m}^2 z_{\rm C}}\right ]~,
\label{dL1}
\eeq
using the Chebyshev polynomial expansion  \cite{capozziello-et-al-2019}, which coefficients of are linear in $\Omega_{\rm m}$.
%
%

We took a set of $714$ SNeIa from the Union2 catalogue \cite{gu17}, devoted to the dark energy search, using data from the Hubble Space Telescope (HST) and combining multiple datasets. 
The catalogue provides the luminosity distance, $d_{\rm L}$, here in parsecs, related to the maximum of the light curve in the B-band 

\vspace*{-0.2cm} 

\beq 
\mu = m - M = 5 \log d_{\rm L} - 5 
\label{mu}
\eeq
where $\mu$ is the distance modulus, $m$ and $M$ are the apparent and absolute magnitudes, respectively. The Union2 catalogue provides also the error on $\mu$, the red-shift $z$ of the host galaxy and its error.  


\begin{table}
\caption{\footnotesize Simulation results for $\Omega_{\rm rad} = \Omega_{\rm k} = \Omega_\Lambda = 0$, $\Omega_{\rm m} = 0.28$. 
 The the error on $k_i$£ is due to the Hierarchical Bayesian Model (HBM) approach, considering different source of uncertainties to render homogeneous the photometric data of the distance modulus from different datasets in Union 2 catalogue. There is no actual error propagation due to the fact that HBM takes into account the error considering it statistically implemented into the model analysis 
The intrinsic standard deviation $\sigma$ for $k_1$ is $3 \times 10^{-6}$. 
The simulation results provide a mean $z_{\rm LSV}$ blue-shift for types 1 and 2 and a $z_{\rm LSV}$ red-shift for type 3.}
\label{tab2}
\newcolumntype{C}[1]{>{\centering\arraybackslash}m{#1}}
\begin{tabular}
{ 
| C{1cm} 
| C{2.4cm} 
| C{2.4cm} 
| C{2.4cm} 
|
} 
\hline
Type
& 1
& 2 
& 3 
\\ \hline
$k_i$  
& $2.8 \times 10^{-5} \pm 3.5 \times 10^{-3}$ 
& $2.6 \times 10^{-5} \pm 2.8 \times 10^{-3}$ 
& $< - 10^{-6} \pm 1.3 \times 10^{-3}$ 
\\ \hline
rms 
& $1.34 \times 10^{-2}$  
& $1.31 \times 10^{-2}$         
& $1.05 \times 10^{-1}$  
\\ \hline
\end{tabular}
\vskip5pt
\end{table}

\begin{figure}[tbp] 
  \centering
  \includegraphics[scale=0.6]{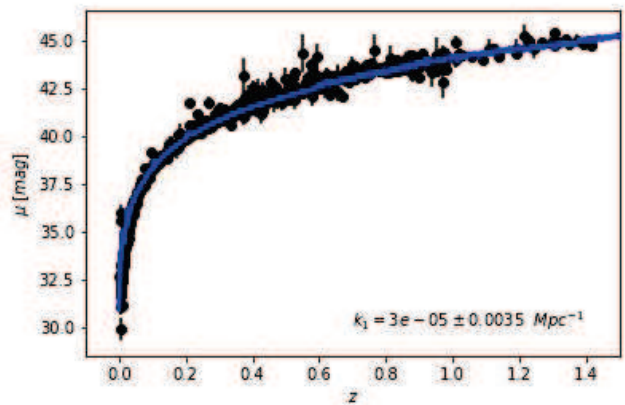}
  \caption{\footnotesize For type 1 of LSV shift, Tab. \ref{tab1}, $\Omega_{\rm rad} = \Omega_{\rm k} = \Omega_\Lambda = 0$ and $\Omega_{\rm m} = 0.28$,  $\mu$, in {\it mag} units, versus $z$, Eq. (\ref{newz}). 
}
  \label{fig:191117_mu(z)}
\end{figure}

\begin{figure}[tbp] 
  \centering
  \includegraphics[scale=0.5]{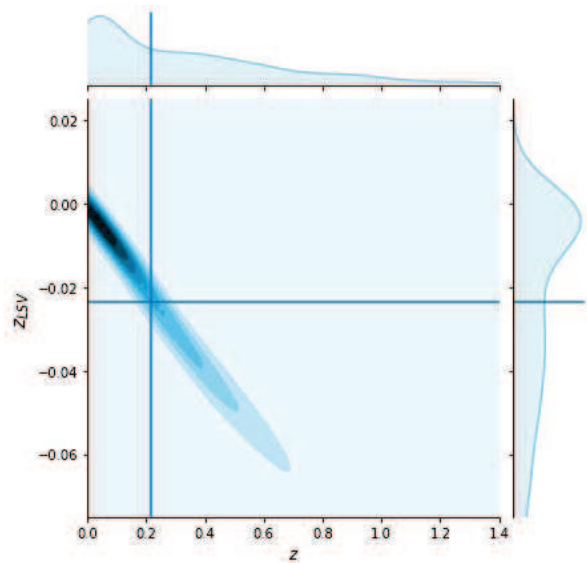}
  \caption{\footnotesize For type 1 of LSV shift, Tab. \ref{tab1}, $\Omega_{\rm rad} = \Omega_{\rm k} = \Omega_\Lambda = 0$ and $\Omega_{\rm m} = 0.28$, the distribution of possible values of $z_{\rm LSV}$ for a given $z$. There are only 73 SNeIa for $z>0.8$.}
  \label{fig:zLSV(z)}
\end{figure}


The comoving distance $r$, computed from the Lema\^itre-Friedman-Robertson-Walker metric (FRWL), is given by 
\cite{peebles-1993,zhang-2013}
\vspace*{-0.2cm} 

\beq
r =\frac{2c}{H}(\sqrt{z_{\rm C}+1}-1)~.
\label{romegam}
\eeq


Our iterative computation starts from the cosmological parameters and compares the final simulated s and the observed value of the distance modulus  

\vspace*{-0.6cm}  

\begin{align}
& \Omega_{\rm k},\Omega_{\rm m} (\Omega_{\Lambda}=0) \rightarrow d_{\rm L} \rightarrow z_{\rm c} \rightarrow r \rightarrow z_{\rm LSV} 
\rightarrow \mu_{\rm s} \stackrel{?}{=} \mu_{\rm o}~, \nonumber
\end{align} 
where the $z_{\rm LSV}$ shift may be positive or negative. 
Herein, we have held to the measured $\mu$ and $z$ while decomposing the latter in two contributions 
$z_{\rm c}$ and $z_{\rm LSV}$. Else, we could have considered the measured $\mu$ as the sum of two contributions 
$\mu(z_{\rm C})$ and $\mu(z_{\rm LSV})$ and looking for an agreement with spectroscopic data.  

{\it Modelling.}
For achieving the best fit, we employed a Hierarchical Bayesian Model (HBM) and determined the entire posterior probability density function of the these parameters. We engaged an HBM to retrieve from the distance distribution of the SNeIa all the possible values of 
$k$, for each of the three types in Tab. \ref{tab1}. We assumed flat priors in the range $[-100,100]$ to avoid heavy constraints in exploring the domains of the parameters.
The Bayesian analysis is very depending upon the choice of the priors, conversely to other kinds, as the Gaussian approaches  \cite{nielsen-guffanti-sarkar-2016, rubin-hayden-2016,dam-heinesen-wiltshire-2017,haridasu-lukovic-dagostino-vittorio-2017,velten-gomes-busti-2018}. The flat priors allow an even probability distribution and thus unbiased estimates.

Unfortunately, the HBM is heavily influenced by the error on the measured values, namely on the  distance modulus leading to a statistical non-informative error on the model parameters. 
From the Gaussian shape of the posterior distribution of $k_i$, it is evinced that a non-zero probability exists for having both positive and negative values of $k_i$, due to the measurement errors. 
We decided to use HBM due to the multiple origin of the distance modulus estimates and differences in data reduction schemes, thus avoiding the influence of some systematic error in the estimates of the parameters. The HBM improves the precision of the estimates of the model parameters, even though the accuracy is rather coarse.  

{\it Results and discussion.}
Table \ref{tab2} provides, for $\Omega_{\rm rad} = \Omega_{\rm k} = \Omega_\Lambda = 0$ and $\Omega_{\rm m} = 0.28$, the $k_i$ values and the associated rms for the three types of $k_i$ associated to $z_{\rm LSV}$. The latter consists of a mean blue-shift for types 1 and 2 and of a red-shift for type 3.  Figure \ref{fig:191117_mu(z)} shows the interpolating curve well fitting $\mu$ and $z$ data, for type 1. Figure \ref{fig:zLSV(z)} shows the distribution of possible values of $z_{\rm LSV}$ for a given $z$, again for type 1. 
We have well interpolating $\mu(z)$ curves also for types 2 and 3 of $z_{\rm LSV}$ shifts.
The rms values of types 1 and 2 are similar, but the frequency variation proportionality to the instantaneous (type 1) rather than emitted (type 2) frequency determines a 'classic' exponential decay or growth. 


The absolute value of $z_{\rm LSV}$ increases with the distance. If blue, the drawback is to explain the photon gaining energy travelling from the source to us, which nevertheless may well occur as a net result of the different contributions of the LSV (vector and tensor) and EM fields along the path. The advantage is a clear understanding of why we measure a lower spectroscopic z, as $z_{\rm LSV}$ is subtracted from the real $z$. If red, the drawback is that $z_{\rm LSV}$ must be comparable or smaller than the observation error  
\cite{palanquedelabrouille-etal-2010,calcino-davis-2017,davis-hinton-howlett-calcino-2019}. The advantage is the correspondence to an intuitive dissipation effect taking place along the photon path. Remarkably, for $\Omega_{\rm m} \geq 0.75$, $k_1$ turns negative and the $z_{\rm LSV}$ becomes a red-shift as in model 3. 


The peculiarity of our approach is that a single mechanism explains all the positions of the SNeIa in the ($\mu, z$) plan, including the outliers. This shift may be red for some SNeIa and blue for others, depending on cosmology parameters, as well as the orientation of the components of the LSV vector or tensor and of the EM fields (host galaxy, intergalactic medium, Milky Way). Evidence of dark energy anisotropy has been shown recently \cite{colin-etal-2019}.
The LSV vacuum energy appears to act on light propagation and it is equivalent to $\Omega_\Lambda \approx 0.7$, rather than igniting an accelerated expansion.

Numerical applications were outlined \cite{helayel-spallicci-2019}. Working them out when considering a SNIa at $z=0.5$, $t_{\rm LB} = 1.57 \times 10^{17}$ s, $f_{i 0}= 3.79 \times 10^{-9}$ V~s~m$^{-2}$, an average $B = 2.75 \times 10^{-9}$ T, the proportionality factor in 
Eq. (\ref{estimate3new}) becomes $5.87 \times 10^{47}$. The parameter $k^{\rm AF}_0$ has a laboratory upper limit of $10^{-10}$m$^{-1}$ but a more stringent, and less favourable for our study, astrophysical upper limit of $5.1 \times 10^{-28}~$m$^{-1}$. In this worst case, it is sufficient that ${\varrho} \geq 7.6 \times 10^{-8}$, to get $z_{\rm LSV}$ in the order of $10\%$ of $z$. 

Where $\Omega_{\rm m}$ was let free, the constraint $\Omega = 1$ leads obviously to huge values of the matter density, as, incidentally,   dark energy emerges at high red-shift.   

{\it Perspectives.}
We have applied to SNeIa a new frequency shift in vacuum occurring to a photon in a LSV scenario, accompanied by an EM field. 
The experimental and observational limits on LSV and magnetic fields are fully compatible with our findings.      

Obviously, the LSV shift is generally applicable and it is not limited to the SNIa case. 
In future work, leaving aside the SME, we will show that a frequency shift is also produced by a generalised non-linear electromagnetism, encompassing the formulations of Born-Infeld and Euler-Kockel-Heisenberg. The found limits on 
$z_{\rm LSV}$ will be applicable to $z_{\rm NL}$ for the non-linear electro-magnetism. 

The XIX century Maxwellian linear electro-magnetism and Einsteinian non-linear gravitation have been well tested. This has not impeded the proposition of alternative formulations of gravity. The lack of experimental proofs on the dark universe and the successes of general relativity prompt to revisit astrophysical observations, largely based on light signals, with non-Maxwellian electro-magnetism, opening the door to radically new interpretations. In future work, we will show that massive photons may fake time dilation for not-moving sources. Such effect, though, is very marginal for SNeIa. 


The $z_{\rm LSV}$ shift is a fraction below 10\% of the measured red-shift and thus below $2 \times 10^{-19} \Delta \nu/\nu$ per m. It would be desirable to test frequency invariance in vacuum \cite{shamirfox1967,wolf1986,wolf1987a,wolf1987b} with a ground or space interferometer.

{\it Acknowledgments.}
FR thanks the Large Synoptic Survey Telescope (LSST) Data Science Fellowship Program, funded by 
 LSST Corporation, the Brinson, Moore and National Science (NSF) Foundations, the Adler Planetarium, and the Center for Interdisciplinary Exploration and Research in Astrophysics (CIERA) at Northwestern University; he acknowledges the NSF CyberTraining Grant \#1829740.
Discussions with L. Barbosa de Lima, M. A. Lopes Capri, V. E. Rodino Lemes, R. de Oliveira Santos (Rio de Janeiro) and S. Capozziello (Napoli) are acknowledged. 

\bibliographystyle{apsrev} 


\end{document}